\documentclass[prb,twocolumn,superscriptaddress,amsmath,amssymb]{revtex4}
\usepackage{natbib}
\usepackage{graphicx}         % Include figure files
\usepackage{dcolumn}         % Align table columns on decimal point
\usepackage{bm}                 % bold math

\begin{document}

\title{Andreev reflection and strongly enhanced magnetoresistance oscillations
in Ga$_x$In$_{1-x}$As/InP heterostructures with superconducting
contacts }

\author{Igor E. Batov}
\email{batov@issp.ac.ru} \altaffiliation[Present address: ] {
Scuola Normale Superiore and NEST CNR-INFM, I-56126 Pisa, Italy.}
\affiliation{Physikalisches Institut III, Universit\"at
Erlangen-N\"urnberg, Erwin-Rommel-Strasse 1, 91058 Erlangen,
Germany} \affiliation{Institute of  Bio- and Nanosystems (IBN-1)
and cni-Center of Nanoelectronic Systems for Information
Technology, Research Centre J\"ulich GmbH, 52425 J\"ulich,
Germany}\affiliation{Institute of Solid State Physics, Russian
Academy of Sciences, 142432 Chernogolovka, Russia}

\author{Thomas Sch\"apers}
\affiliation{Institute of  Bio- and Nanosystems (IBN-1) and
cni-Center of Nanoelectronic Systems for Information Technology,
Research Centre J\"ulich GmbH, 52425 J\"ulich, Germany}

\author{Nikolai M. Chtchelkatchev}
\affiliation{L. D. Landau Institute for Theoretical Physics, Russian
Academy of Sciences, 117940 Moscow, Russia}

\author{Hilde Hardtdegen}
\affiliation{Institute of  Bio- and Nanosystems (IBN-1) and
cni-Center of Nanoelectronic Systems for Information Technology,
Research Centre J\"ulich GmbH, 52425 J\"ulich, Germany}

\author{Alexey V. Ustinov}
\affiliation{Physikalisches Institut III, Universit\"at
Erlangen-N\"urnberg, Erwin-Rommel-Strasse 1, 91058 Erlangen,
Germany}

\date{\today}% It is always \today, today,

\hyphenation{GaInAs}

\begin{abstract}
We study the magnetotransport in small hybrid junctions formed by
high-mobility Ga$_x$In$_{1-x}$As/InP heterostructures coupled to
superconducting (S) and normal metal (N) terminals. Highly
transmissive superconducting  contacts to a two-dimensional
electron gas (2DEG) located in a Ga$_x$In$_{1-x}$As/InP
heterostructure are realized by using a Au/NbN layer system. The
magnetoresistance of the S/2DEG/N structures is studied as a
function of dc bias current and temperature. At bias currents
below a critical value, the resistance of the S/2DEG/N structures
develops a strong oscillatory dependence on the magnetic field,
with an amplitude of the oscillations considerably larger than
that of the reference N/2DEG/N structures. The experimental
results are qualitatively explained by taking Andreev reflection
in high magnetic fields into account.
\end{abstract}

\maketitle

\section{Introduction}

Mesoscopic systems consisting of superconductor/semiconductor
hybrid structures have attracted considerable attention in recent
years.\cite{Wees97,Kroemer99a,Schaepers01a} The carrier transport
in superconductor/normal metal or superconductor/semiconductor
structures can be described in the framework of Andreev
reflection.\cite{Andreev64} During an Andreev reflection process
an electron that travels from the semiconductor on a
superconductor/semiconductor interface is retroreflected as a
hole. Simultaneously, a Cooper pair is created on the
superconductor side. A number of interesting phenomena based on
Andreev reflection had been studied in the past, e.g. gate-control
of a Josephson supercurrent,\cite{Takayanagi85,Doh05}
superconducting quantum point contacts,\cite{Takayanagi95d}
control of the supercurrent by hot carrier
injection,\cite{Morpurgo98,Schaepers98} and supercurrent reversal
in a quantum dot.\cite{VanDam06}

A two-dimensional electron gas (2DEG) in a semiconductor offers
the advantage of ballistic transport in the semiconductor part. A
fascinating regime occurs in high magnetic fields as soon as the
transport across the superconductor/2DEG is governed by the Landau
quantization in the
2DEG.\cite{Ma93,Fisher94,Takagaki98,Hoppe00,Zuelicke01,Giazotto05}
Microscopic
calculations\cite{Takagaki98,Hoppe00,Zuelicke01,Giazotto05}
revealed conductance oscillations in S/2DEG junctions as a
function of magnetic field. It was theoretically shown by Hoppe
{\it et al.}\cite{Hoppe00} that the current flow along the S/2DEG
interface can be described in the framework of Andreev bound
states formed by electron and hole edge state excitations. At
lower magnetic fields one can view this process in a semiclassical
picture, in which carrier propagation is maintained by skipping
orbits of electrons and holes along the
interface.\cite{Asano00,Asano00a,Chtchelkatchev01,Asano00b,Chtchelkatchev07}
In mesoscopic S/2DEG contacts where the phase coherence is
maintained during the quasiparticle propagation, the interference
between electrons and Andreev-reflected holes can lead to the
magnetoconductance oscillations which are based on a
Aharonov--Bohm type
effect.\cite{Asano00,Asano00a,Chtchelkatchev01,Asano00b} The
semiclassical theory of the charge transport through the S-2DEG
interface at large filling factors was developed in
Refs.~\onlinecite{Chtchelkatchev01,Chtchelkatchev07}. Apart from
the orbital effects, a magnetic field can also be employed to
induce Zeeman energy splitting in the 2DEG. This opens up the
possibility to study spin-related effects in combination with
Andreev
reflection.\cite{Bezuglyi02,Tkachov05,Giazotto05,Frustaglia05}

From experimental point of view, it is challenging to fabricate
highly transmissive superconducting contacts to a 2DEG using
superconductors with high critical magnetic
fields.\cite{Takayanagi98,Uhlisch00,Moore99} Recently, Eroms {\it
et al.}\cite{Eroms05} found enhanced oscillations in the
magnetoresistance of a Nb/InAs structure for magnetic fields below
the critical field of Nb.

In this work, we report on the magnetotransport across a NbN/Au/2DEG
interface. The choice of the NbN/Au system was motivated by our
previous studies, where an Au interlayer helped achieving a high
S/2DEG interface transparency while maintaining a high critical
field of the superconductor.\cite{Batov04} Complementary to the work
of Eroms {\it et al.},\cite{Eroms05} we observe a suppression of
enhanced oscillations in the magnetoresistance when a dc bias
current across the junction exceeds a critical value or the
temperature is increased above a critical temperature.\cite{Batov04}
We compare our measurements of the NbN/Au/2DEG structures to that of
similar structures with normal metal electrodes connected to the
2DEG. Our interpretation of the experimental findings is based on
recent theoretical models describing Andreev reflection across a
S/2DEG interface in the presence of a magnetic field.

\section{Experimental}

\begin{figure}
\includegraphics[width=0.65\columnwidth]{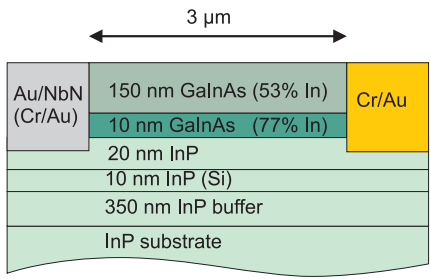}
\caption{(Color online) Schematics of the sample cross section.
The size of the mesa is $3 \times 3$~$\mu$m$^2$. For the first
type of structures a superconducting Au/NbN electrode and a
normally conducting Cr/Au electrode face each other. For the
second sample type two Cr/Au electrodes were used. \label{Fig1}}
\end{figure}

The strained Ga$_x$In$_{1-x}$As/InP heterostructure was grown on a
semi-insulating InP substrate by using metal organic vapor phase
epitaxy (MOVPE). Figure~\ref{Fig1} shows the corresponding layer
sequence. The 2DEG is located in the strained
Ga$_{0.23}$In$_{0.77}$As layer. From Shubnikov--de Haas measurements
on Hall bar samples a carrier concentration of $n=6.3 \times
10^{11}$~cm$^{-2}$ and a mobility of $\mu=250,000$~cm$^2$/Vs was
extracted at 0.3~K. Analysis of the temperature-dependent
Shubnikov--de Haas oscillations yielded an effective electron mass
$m^*=0.036~m_{e}$, which is in good agreement with previously
reported results.~\cite{Schaepers04a} Based on the values given
above, a transport mean free path $l_{\rm tr}$ of 3.3~$\mu$m and a
Fermi energy $E_{F}$ of 42~meV were determined.

We used a three-step electron beam lithography process to fabricate
the samples. First, the mesa was defined by CH$_4$/H$_2$ reactive
ion etching using a Ti layer as an etching mask. The etching depth
of 170~nm was well below the depth of the Ga$_{0.23}$In$_{0.77}$As
channel layer. In the second step the superconducting electrodes (S)
contacting the 2DEG at the mesa sidewalls were defined by electron
beam lithography. We used Ar plasma cleaning to remove residual
atoms on the surface. Subsequently, a 10 nm thin Au interlayer
followed by a 100~nm thick NbN layer were deposited \emph{in-situ}
by dc magnetron sputtering. After the lift-off the sample was
annealed at a temperature of $400^\circ$C for 10~sec. The Au
interlayer and the annealing were introduced to improve the
interface transparency.\cite{Batov04} By the final electron beam
lithography step, the normally conductive Cr/Au electrodes (N)
(5nm/100nm) were deposited by electron beam evaporation.
Figure~\ref{Fig1} shows a sketch of the sample cross section. The
size of the 2DEG mesa was $3 \times 3$~$\mu$m$^2$. The S/2DEG
interface length $L$ in our samples was 3 $\mu$m. Two types of
structures were prepared. In the first type (S/2DEG/N) a
superconducting Au/NbN electrode and a normal conductive electrode
were facing each other, whereas for the second type (N/2DEG/N) a
normal conductive material (Cr/Au) was used for both electrodes.

All measurements were performed in a He-3 cryostat in a
two-terminal configuration. The sample resistance was measured by
employing the current-driven lock-in technique with an ac
excitation current of 10~nA. In order to add an additional dc bias
voltage across the sample, a dc-current $I_{dc}$ was superimposed
for some measurements.

\section{Results and Discussion}

\begin{figure}
\includegraphics[width=\columnwidth]{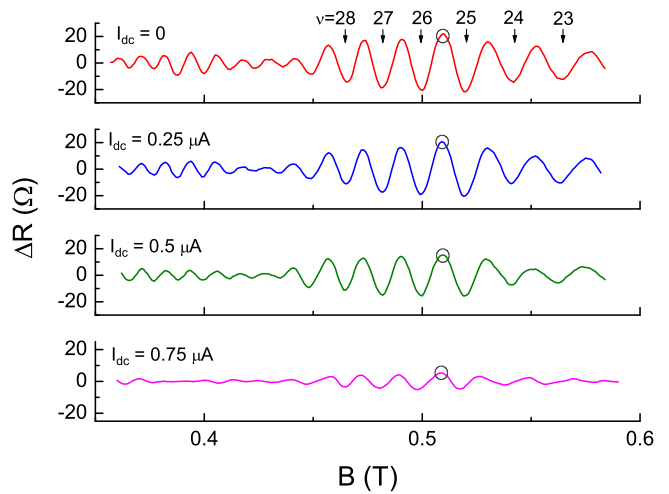}
\caption{(Color online) Magnetoresistance oscillations $\Delta R$
of the S/2DEG/N sample for various dc bias currents $I_{dc}$: 0,
0.25, 0.5, 0.75~$\mu$A at a temperature of 0.5~K. The oscillation
amplitude was extracted for Fig.~3 at the magnetic field value of
0.51~T indicated by a circle. The filling factors $\nu$ are
indicated by arrows. \label{Fig2}}
\end{figure}

At the magnetic fields investigated ($B<0.6$~T), the two-terminal
measurements showed a positive magnetoresistance with superimposed
oscillations in all samples. \footnote{The beating pattern of the
magnetoresistance can be attributed to the presence of the Rashba
effect, see e.g. Th. Sch\"apers, G. Engels, J. Lange, \mbox{Th.}
Klocke, M. Hollfelder, and H. L\"uth, J. Appl. Phys., {\bf 83},
4324 (1998)} In Fig.~\ref{Fig2} we display data obtained for an
S/2DEG/N structure after subtracting the slowly varying positive
magnetoresistance background. \footnote{The background resistance
contribution was gained by fitting a polynomial curve to selected
measurements points being midway within the oscillating resistance
contribution.}

\begin{figure}
\includegraphics[width=\columnwidth]{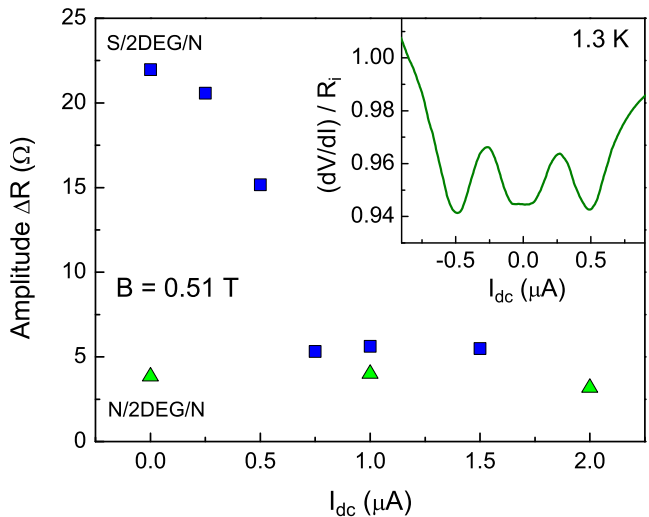}
\caption{(Color online) Oscillation amplitude $\Delta R$ at
$B=0.51$~T as a function of the dc bias current $I_{dc}$ for the
S/2DEG/N sample (squares) and for the N/2DEG/N structure
(triangles). The inset shows the normalized differential
resistance $(dV/dI)/R_i$ of the S/2DEG/N structure as a function
of $I_{dc}$. \label{Fig3}}
\end{figure}

At low dc bias currents the S/2DEG/N structure reveals clear
resistance oscillations as a function of magnetic field.
Figure~\ref{Fig3} shows the dependence of the amplitude of the
resistance oscillations on the dc bias current at 0.51~T extracted
from the data plotted in Fig.~\ref{Fig2}. For comparison, the
corresponding oscillation amplitudes for the reference N/2DEG/N
sample are also shown in Fig.~\ref{Fig3}. It can be seen, that
within the bias current range $I_{dc} \lesssim 0.5$~$\mu$A the
oscillation amplitude in the S/2DEG/N structure is substantially
enhanced over that of the N/2DEG/N structure. At zero dc bias
current the oscillation amplitude in the S/2DEG/N structure are
larger by a factor of about 5.

The magnetoresistance oscillation amplitudes in the S/2DEG/N
samples show two distinctly different regimes as a function of
$I_{dc}$. As can be seen in Fig.~\ref{Fig3}, in the range $0 \leq
I_{dc} < 0.75 $~$\mu$A the amplitude of the magnetoresistance
oscillations $\Delta R$ decreases monotonously with increasing
bias current, comprising a sharp drop for dc currents exceeding
0.5~$\mu$A. At currents $I_{dc} \geq 0.75$~$\mu$A, the amplitude
of the resistance oscillations shows only a very weak bias current
dependence. In strong contrast, in the N/2DEG/N reference
structures, the magnetoresistance oscillation amplitude depends
only weakly on the dc bias current in the whole range from zero up
to 2~$\mu$A, as shown in Fig.~\ref{Fig3}.

As it can be seen in the inset of Fig.~\ref{Fig3}, at low
temperatures the differential resistance $(dV/dI)/R_i$ of S/2DEG/N
sample, normalized to the resistance $R_i$ at 1.5~$\mu$A, shows a
decrease within the range of dc bias currents of $\pm 0.8$~$\mu$A.
In order to consider the contribution of the S/2DEG interface
only, we subtracted the resistance of the 2DEG/N interface deduced
from the N/2DEG/N reference structure. Our previous measurements
of S/2DEG single junctions prepared in the same processing run
revealed that the decrease in the differential resistance is
related to the superconducting energy gap.\cite{Batov04} This
suggests that the enhanced magnetoresistance oscillations detected
at low dc bias currents are most likely due to the
Andreev-reflection contribution to the interface conductance.

The decrease of the differential resistance at low dc bias
currents indicates that the barrier at the S/2DEG is relatively
low. A transmission coefficient $T_N=0.74$ was estimated from the
ratio of the resistances at zero bias and large bias currents,
following the Blonder-Tinkham-Klapwijk model.\cite{Blonder82} The
high transmission probability results from the Au layer introduced
between the NbN layer and the 2DEG.\cite{Batov04} The finite
interface barrier can be attributed partly to the Fermi velocity
mismatch between the metallic layer and the 2DEG and partly to
contamination at the interface. The specific shape of the
$(dV/dI)/R_i-I_{dc}$ characteristics can be associated with the
presence of the Au interlayer. Due to the proximity effect between
the superconducting NbN layer and the Au layer, a gap in the
density of states is induced in the normal conducting Au film
resulting in the maxima in the differential resistance
$(dV/dI)/R_i$ observed at approximately $\pm 0.27
\mu$A.\cite{Batov04}

\begin{figure}
\includegraphics[width=\columnwidth]{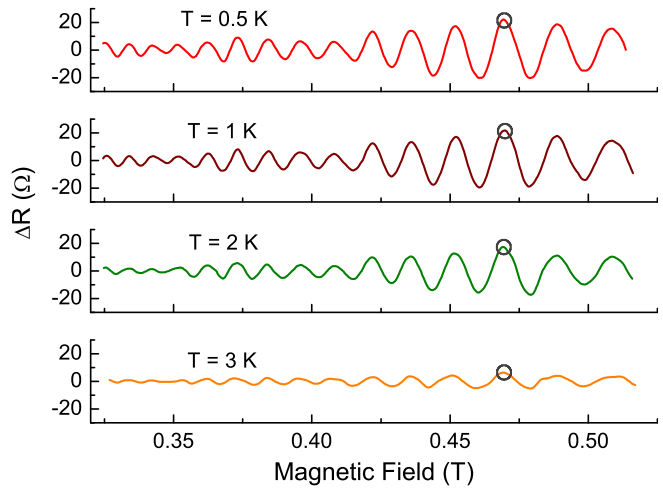}
\caption{(Color online) Magnetoresistance oscillations $\Delta R$
of the S/2DEG/N sample at $I_{dc}=0$ for various temperatures:
0.5, 1.0, 2.0, 3.0~K. \label{Fig4}}
\end{figure}

Figure~\ref{Fig4} shows magnetoresistance oscillations of the
S/2DEG/N structure at different temperatures in the 0.5--3~K range.
The data are taken at zero dc bias current after subtracting the
background resistance. The oscillation amplitudes are found to
decrease with increasing temperature. Our experiments indicate that
the magnetoresistance oscillations are very sensitive to
temperature. With the assumption that the conventional effective
mass approach is applicable,\cite{Hang01} we attempted to determine
the effective mass from the temperature dependence of the
oscillation amplitude. However, in consistence with the results of
Eroms \emph{et al.}\cite{Eroms05} the fit was poor. Below we will
show that the experimental temperature dependence of the oscillation
amplitudes can be explained within the model of the phase-coherent
transport of electrons and Andreev reflected holes in S/2DEG
junctions in a ballistic regime.
\cite{Asano00,Asano00a,Chtchelkatchev01,Asano00b}

\begin{figure}
\includegraphics[width=1.0\columnwidth]{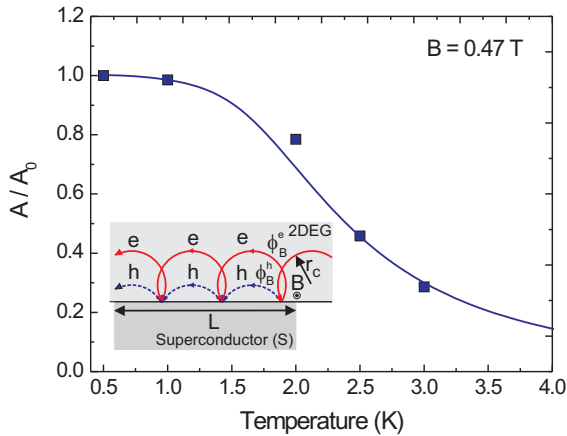}
\caption{(Color online) Normalized amplitude $A/A_0$ as a function
of temperature at $B=0.47$~T (squares). Here, $A_0$ is the
amplitude at $T=0.5$~K. Solid line represents the calculated
amplitude according to Ref.~\onlinecite{Chtchelkatchev01}. The
inset shows a schematics of the Andreev reflection process at
S/2DEG interface with a magnetic field applied perpendicularly to
the plane of the 2DEG. The electrons and holes acquire a phase
shift $\phi_{B}^e$ and $\phi_{B}^h$ between two successive Andreev
reflections. The quantities, $r_c$ and $L$ denote the cyclotron
radius and the length of the S/2DEG interface, respectively.
\label{Fig5}}
\end{figure}

The magnetoconductance of superconductor/2DEG junctions in high
magnetic fields was theoretically studied in
Refs.~\onlinecite{Asano00,Asano00a,Chtchelkatchev01,Asano00b,Takagaki98,Hoppe00,Zuelicke01,Giazotto05}.
In Refs.~\onlinecite{Asano00,Asano00a,Chtchelkatchev01,Asano00b} it
has been shown that the magnetoconductance oscillations appear in
the high-field regime in a ballistic junction when the Andreev
reflection is not perfect at the interface and the diameter of the
cyclotron motion of quasiparticles is smaller than the width of the
junction. The mechanism of the novel magnetoconductance oscillations
has been revealed in Refs.~\onlinecite{Asano00a,Asano00b}, based on
the phenomenology of the Aharonov--Bohm type interference effect,
and can be explained using a semiclassical description of a charge
transport in the S/2DEG junction. Figure~\ref{Fig5}~(inset)
illustrates a semiclassical picture for Andreev reflection process
at a finite magnetic field.
In case of a barrier at the S/2DEG interface, an electron
impinging at the interface is reflected to a certain probability
as an electron or as a hole. The magnetic field forces the
quasiparticles to circular motion. Due to the opposite effective
mass and the inverse charge, the electron and hole orbits do have
the same chirality, as shown in Fig.~\ref{Fig5}~(inset). Thus,
both quasiparticles propagate in the same direction along the
interface. Depending on whether an electron or a hole is Andreev
reflected at the interface, a Cooper pair is formed or removed
from the superconductor, respectively, resulting in a net current
across the S/DEG interface. The electron (hole) wave acquires a
phase shift $\phi_{B}^e$ ($\phi_{B}^h$) on the path between two
scattering processes at the interface, due to the magnetic field,
during the circular motion in the 2DEG. It has been shown that the
phase difference between the pair of the waves $\phi_{B}^e -
\phi_{B}^h$ is independent on the incident angle of the electron
and proportional to the magnetic flux encircled by the single
complete cyclotron orbit.\cite{Asano00a,Asano00b} This results in
the Aharonov--Bohm type interference of quasiparticles at the
interface.\cite{Asano00,Asano00a,Chtchelkatchev01,Asano00b} As a
consequence, the zero-bias conductance oscillates as a function of
magnetic flux encircled by the cyclotron orbit in units of
$\phi_{0} = h/e$. The magnetoconductance oscillations are periodic
as a function of the inverse magnetic
field.\cite{Asano00a,Asano00b} In order to establish periodic
oscillations, the length $L$ of the S/2DEG interface must be
larger than the cyclotron diameter $2r_c=2\hbar\sqrt{2\pi n}/eB$
and smaller than the transport mean free path $l_{\rm tr}$. This
ensures that the quasiparticles impinge at the interface at least
twice, in order to allow for interference. In our case, the
largest possible cyclotron diameter is 880~nm corresponding to the
lowest magnetic field of 0.3~T considered here. In addition, the
length of the S/2DEG interface $L$ is smaller than $l_{\rm tr}$.
Thus, both conditions are fulfilled. Note that the theoretical
analysis \cite{Asano00,Asano00a,Chtchelkatchev01,Asano00b}
presented above is not valid at very high magnetic fields when the
filling factors $\nu$ are in the order of unity. In this case the
electron and hole propagation can be described in terms of Andreev
edge states.\cite{Hoppe00,Zuelicke01,Giazotto05} Similar to the
results reported in
Refs.~\onlinecite{Asano00,Asano00a,Chtchelkatchev01,Asano00b,Takagaki98},
analytical calculations within the edge channels picture showed
the pronounced magnetoconductance oscillations periodic in $1/B$
for the case of a finite barrier at the interface or a mismatch of
the Fermi velocity.\cite{Hoppe00,Zuelicke01,Giazotto05} The
amplitude of the magnetoconductance oscillations as a function of
the dc bias voltage was studied numerically in
Refs.~\onlinecite{Asano00b,Takagaki98}. It has been shown that the
pronounced magnetoconductance oscillations can also be seen at the
finite bias voltage $V<\Delta_0/e$. The oscillations are
suppressed at $V>\Delta_0/e$, since the amplitude of the Andreev
reflection is significantly reduced for $eV/\Delta_0>1$.

In accordance to the theoretical predictions cited above, we found
pronounced oscillations in the magnetoresistance of the S/2DEG/N
structure. However, a similar oscillation period in $1/B$
(Shubnikov-de Haas oscillations) is expected for the
magnetoresistance of the 2DEG as well. We therefore have to make
sure that the enhancement of oscillations observed in our
experiment can indeed be attributed to Andreev refection at the
S/2DEG interface. The first evidence is that a considerably larger
oscillation amplitude is found for the S/2DEG/N sample compared to
the N/2DEG/N structure (see Fig.~\ref{Fig3}). An enhanced
oscillation amplitude for the magnetoresistance of a S/DEG
structure at bias voltages $V$ less than $\Delta_0/e$ was
theoretically predicted in case of a finite interface
barrier.~\cite{Asano00b,Takagaki98} As can be seen in
Fig.~\ref{Fig3}, the large-amplitude oscillations are preserved up
to dc bias currents $I_{dc} \sim0.75$~$\mu$A. At bias current
$I_{dc}$ above this value the bias voltage $V$ exceeds the voltage
$\Delta_0/e$ related to the superconducting gap energy $\Delta_0$,
as indicated by the measurement of the differential resistance. A
similar behavior was found in the calculations by Asano {\it et
al.}\cite{Asano00b} and Takagaki.\cite{Takagaki98} There, an
abrupt decrease of the oscillation amplitude was observed at
$eV/\Delta_0>1$. The experimentally observed dependence of the
oscillation on the bias current is in strong contrast to the
findings regarding the N/2DEG/N structure, where the oscillation
amplitude remains constant for the entire range of $I_{dc}$.

Our interpretation is supported further by the measurements of the
magnetoresistance as a function of temperature. Here, a strong
decrease of the oscillation amplitude with increasing temperature
was observed at temperatures below the superconducting transition
temperature $T_c$ in our samples. Based on the semiclassical model
for the current transport in a ballistic S/2DEG junction,
\cite{Chtchelkatchev01} we have estimated the temperature
dependence of the magnetoresistance oscillation amplitudes. In the
calculations, the normal and Andreev reflection coefficients are
approximated by the BTK model.\cite{Blonder82} The superconducting
energy gap $\Delta(T)$ in the superconductor is assumed to follow
the BCS temperature dependence. \cite{Schrieffer99,Tinkham96} The
differential conductance at zero bias $dI/dV(T)$ is evaluated by
integration over energy of the spectral
conductance\cite{Chtchelkatchev01,Chtchelkatchev07} multiplied by
the energy derivative of the Fermi distribution function. In
Fig.~\ref{Fig5} we show the results of our calculations of the
temperature dependence of the oscillation amplitude. In the
simulations, for the semiconductor region we used parameters
characteristic of our InGaAs/InP heterostructures. The interface
barrier-strength parameter $Z$ was estimated from the experimental
data, and the superconducting energy gap parameter $\Delta_0$ was
chosen to adjust the experimental temperature dependence of the
oscillation amplitude. The calculated results are found to be in
good agreement with the experimental data for values of the
$\Delta_0$ parameter close to the superconducting energy gap in
the sample measured by the differential resistance versus bias
voltage characteristics. Thus, the model of the phase-coherent
transport of carriers at the S/2DEG interface in strong magnetic
fields \cite{Asano00,Asano00a,Chtchelkatchev01,Asano00b} appeared
to be consistent with our experimental data. In
Ref.~\onlinecite{Batov04} we have analyzed finite temperature
zero-field current-voltage characteristics of NbN/Au/GaInAs-InP
junctions within a model based on the quasiclassical
Green-function
approach.\cite{Neurohr96,Golubov96,Belzig99,Golubov04} At present,
however, the theoretical description of the conductance
oscillations at an S/2DEG interface in a magnetic field based on
this approach has not yet been developed.

A similar temperature behavior of the magnetoresistance was found by
Eroms {\it et al.}\cite{Eroms05} They attributed the enhanced
oscillation amplitude to the higher backscattering contribution in
the 2DEG due to the combined occupation of the edge channels by
electrons and holes in case of Andreev reflection at the interface.
Giazotto \emph{et al.}\cite{Giazotto05} studied theoretically the
effect of Zeeman splitting on the Andreev reflection at the
superconductor/2DEG interface. It is predicted that the effect of
Zeeman splitting should be visible as a double-step feature in the
conductance of transparent S/2DEG interface. However, at magnetic
fields investigated here these corrections are small and thus the
effect of the Zeeman splitting could not be resolved in our
experiments.

\section{Conclusions}

In summary, we have investigated the magnetotransport in S/2DEG/N
structures at various dc bias currents and temperatures. We have
found that the amplitude of oscillations of the magnetoresistance
is considerably enhanced at low bias currents. The observed
behavior is interpreted using the framework of phase-coherent
Andreev reflection in the presence of a magnetic field.

\begin{acknowledgments}
The authors thank U.~Z\"ulicke (Massey University, New Zealand),
A. A. Golubov (Twente University, The Netherlands), and I. S.
Burmistrov (Landau Institute, Moscow) for fruitful discussions.
This work was supported by the Deutsche Forschungsgemeinschaft
(DFG).
\end{acknowledgments}

\end{document}